# JUICY OR DRY? A COMPARATIVE STUDY OF USER ENGAGEMENT AND INFORMATION RETENTION IN INTERACTIVE INFOGRAPHICS

Bruno Campos[1]

[1]Department of Design, MacEwan University, Edmonton, Canada
camposb@macewan.ca

## ABSTRACT

*This study compares the impact of "juiciness" on user engagement and short-term information retention in interactive infographics. Juicy designs generally showed a slight advantage in overall user engagement scores compared to dry designs. Specifically, the juicy version of the Burcalories infographic had the highest engagement score. However, the differences in engagement were often small. Regarding information retention, the results were mixed. The juicy versions of The Daily Routines of Famous Creative People and The Main Chakras infographics showed marginally better average recall and more participants with higher recall. Conversely, the dry version of Burcalories led to more correct answers in multiple-choice questions. The study suggests that while juicy design elements can enhance user engagement and, in some cases, short-term information retention, their effectiveness depends on careful implementation. Excessive juiciness could be overwhelming or distracting, while well-implemented juicy elements contributed to a more entertaining experience. The findings emphasize the importance of balancing engaging feedback with clarity and usability.*

## KEYWORDS

*Infographics, Juiciness, Interactive, Engagement*

## 1. INTRODUCTION

Interactive infographics aim to leverage the power of visual representation and interactive features to deliver information effectively, tell compelling stories with data, and potentially influence attitudes and behaviours in digital environments (Greussing and Boomgaarden, 2021; Zwinger and Zeiller, 2016). Past research has identified and suggested potential opportunities for techniques and exploration regarding interactivity within infographics, such as the integration of game-like mechanics to increase user engagement and exploration (Diakopoulos et al, 2011).

Within games, a technique that has been widely explored to increase user engagement through abundant feedback with high-level aesthetic appeal is the concept of Juiciness (Hicks, 2020; Durmanova, 2022; Kao, 2020). Juiciness has also been explored in non-gaming settings, often paired with gamification to enhance user engagement and motivation. (Durmanova, 2022). Previous research has looked for the presence of juicy elements outside of game environments (Campos, 2025), encountering potential to further explore juiciness in other interactive mediums, such as interactive infographics.

While interactive visualizations often succeed in their affective function by generating situational interest, enjoyment, and appeal, the evidence regarding their cognitive function, such as processing, storing, and recalling content, is less clear (Greussing and Boomgaarden, 2021).

This research proposes a comparative study to understand the impacts of Juiciness regarding user engagement and short-term information retention in the context of interactive infographics. To do so, it is proposed the following research questions:

**RQ: How does the incorporation of juicy design elements in interactive infographics influence user engagement and information retention compared to dry/regular designs?**

To support this research question, the following research objectives and hypothesis are presented:

Research Objectives:

1. To assess the impact of juicy design elements on user engagement in interactive infographics.
2. To evaluate whether juicy design elements improve the short-term retention of information presented in interactive infographics.

Hypothesis:

H1: Interactive infographics with juicy design elements will result in higher user engagement scores than dry/regular infographics.
H2: Users interacting with juicy infographics will exhibit better short-term information rates retention compared to those interacting with dry/regular infographics.

This research is expected to contribute both academically and practically. Academically, it merges the emerging field of Juiciness with the well-established area of interactive infographics and data visualizations. Practically, it offers insights for practitioners who may integrate Juicy elements into their design process and explore their potential with users.

The paper is structured as follows: (2) a Literature Review on Interactive Infographics, Juiciness, and Information Retention; (3) a Methodology section detailing the infographics development process, highlighting differences between Dry and Juicy versions, and providing an overview of participants and procedures; (4) presentation of the Results; (5) a Discussion that interprets the findings and addresses limitations and suggestions for future research; and (6) a Conclusion offering final thoughts.

## 2. LITERATURE REVIEW

### 2.1. Interactive Infographics and User Engagement

Infographics are visual representations of information and data. (Gonzalez, 2018) They integrate images, text, numbers, visual design, and Web technology to provide a visual explanation of complex phenomena. (Greussing and Boomgaarden, 2021) The goal of infographics is to transform unstructured information into graphical compositions that are both easy to understand and visually appealing, aiming to facilitate the understanding of facts, processes, and data. (Zwinger and Zeiller, 2016; Won, 2018; Costa Pinto, 2017) They can combine data visualizations, illustrations, text, and images together into a format that tells a story and communicates insights in a visual form. (Gonzalez, 2018)

Zwinger (et al, 2017) state interactive infographics built upon the previous foundation by offering the user at least one option to modify the form or content of the graphic in real time. The same authors say that they are visual representation of information that integrates several modes (at least two), such as image/video, spoken or written text, audio, layout, and others, to a coherent ensemble that offers at least one option of control to the user. These controls can include start/stop buttons, forward/backward buttons, menu items, timelines, filters, or input boxes. Interactive infographics can also supply all the information found in static infographics along with multimedia resources and include controls that allow readers to make changes to the data and visualizations. (Dehghani et at, 2020; Yldirim, 2017; Afify, 2018)

According to Lalmas (*et al*, 2015) and O'Brien (*et al*, 2018), user engagement (UE) is a multifaceted concept that has garnered increasing interest in human-computer interaction (HCI) and various other fields. The authors characterize user engagement as the depth of an actor's investment when interacting with a digital system. It is considered a quality of user experience (UX), focusing on the positive aspects of interacting with an online application and, in particular, the desire to use that application longer and repeatedly. (Lalmas *et al*, 2015) The same author proposes that, in a broad sense, user engagement encompasses a user's emotional connection, mental focus, and active participation with a technology over time, driven by their needs and the qualities of the system.

As previously noted, interactive infographics are defined by their ability to offer users at least one option to modify the form or content of the graphic in real time, allowing for engagement with the information through functional tools available on the interface. These interactive features can significantly impact user engagement in various ways. Media practitioners believe that interactive infographics are effective in attracting audiences' initial attention, as they offer the ability to interact with an image in a playful way, not only catching users' attention but also serving as a heuristic cue that invokes conscious acknowledgments of the novelty and aesthetic pleasure of the digital environment. (Greussing and Boomgaarden, 2021; Coelho and Mueller, 2020) Chart animations, which are a form of interactive visualization, are considered an effective means of attracting people's attention and maintaining their engagement. (Ge *et al*, 2020) By allowing users to control the presentation of information, interactive infographics can enhance readers' concentration and motivation to engage with information. (Burnett *et al*, 2019)

Still, it is important to point out that some research suggests potential downsides of incorporating interaction to infographics in relation to engagement. Greussing and Boomgaarden (2021) mention that interactivity can place considerable demands on news consumers' limited cognitive resources, potentially leading users to focus more on the interaction task itself than on the underlying information. The authors say that if an interactive visualization is considered complex or confusing, users may not be able or willing to thoroughly process all information available and may be distracted from enjoyable consumption. Nevertheless, interactive infographics have the potential to significantly enhance user engagement through their novelty, aesthetic appeal, provision of user control, and ability to convey information in an engaging manner. (Won, 2017)

## 2.2. Juiciness and User Experience

The concept of "juiciness" in game design refers to the use of exaggerated audio-visual and haptic feedback to enhance player experience. The term "juice" was first introduced around 2005, described as a "wet little term for constant and bountiful user feedback" (Kucic, 2005). A juicy game element will exhibit noticeable reactions like bouncing, wiggling, squirting, and making noise upon interaction. (Kao *et al*, 2024; Durmanova, 2022) Game developers define and utilize "juiciness" in game design as a key element for enhancing player experience through abundant and satisfying feedback (Johansen and Cook, 2021; Hicks, 2020). While the term itself originated informally, it has become a widely recognized concept in the game industry and is increasingly studied in academia. (Hicks *et al*, 2024; Fabre *et al*, 2024)

Juicy design significantly affects user experience in various ways, generally aiming to create a more engaging and satisfying interaction. (Hicks et al, 2019) Previous research detail studies investigating how juicy elements impact player perception, engagement, and even physiological responses like attention and disengagement. Developers utilize juiciness to foster a positive emotional response, a feeling of reward and satisfaction, and an overall enjoyment of being within the game world, contributing to creating a more immersive and enjoyable experience for players. (Durmanova, 2022; Atanasov, 2013) Juiciness is considered an "interaction aesthetic" or a positively beautiful quality of interaction at the "motor level", being about the immediate, moment-to-moment feedback loop between player input and game response. (Kao et al, 2024)

Although research has found out that Juiciness overall implies and generates a positive experience within users, the impact of Juiciness on user experience is not always positive. Too much "juiciness" can be overwhelming and may reduce the player's intrinsic motivation to keep playing. (Luz et al, 2024) It can also be irritating to users due to excessive graphics and sound effects. (Kao, 2020) Too little "juiciness" can result in a less engaging and appealing experience. Participants from a study noted that non-juicy versions felt like something was either missing, or incomplete. (Hicks, 2020) There's also evidence of a "Goldilocks effect" - the human tendency to prefer options that are neither too extreme nor too moderate, but fall within an optimal, balanced range (Kao et al, 2024). The same author found that medium and high juiciness conditions outperformed low and extreme juiciness across measures like interest/enjoyment and presence/immersion.

These quotes provide an overview about Juiciness and its impacts to user experience. It is believed that there's potential room to further explore its application on interactive infographics, verifying to what extent juiciness would affect user engagement and short-term information retention from infographics content.

## 2.3. Information Retention and Interactive Visualizations

Information retention can be defined as the ability to remember or recall information that has been previously presented, often assessed by measures such as accurate recall or recognition of the material. (Savoy et al, 2009; Fischer et al, 2023; Schechinger, 2023) Levels of information retention can be distinguished between long-term retention, referring to the ability to remember information over a more extended period, ranging from days to weeks, months, or even years (O'Day, 2007); and short-term retention, being closely related to working memory, which has a limited capacity, and a short duration, approximately 30 seconds (Zeglen and Rosendale, 2018)

According to Kizilcec (et al, 2014), a common method to verify short-term retention is to administer recall tests very soon after the presentation of information. These can be multiple-choice questions, select-all-that-apply items, open-ended questions for free recall, amongst others. Some of those techniques are applied and further discussed during the methodology section.

Research from Fischer (et al, 2023) and Schechinger (2023) state that information retention is significantly related to visualizations, as they can influence how information is encoded, stored, and retrieved in memory. The authors mention that by presenting quantitative data in visual formats like images, signs, maps, graphs, and charts, communicators can deliver information quickly, helping individuals to encode information more effectively. Visualizations, such as graphics in lectures or data visualizations in infographics, can facilitate the initial encoding of information into working memory. (Shepard and Teghtsoonian, 1961; Fischer et al, 2023) By presenting information visually, it can be captured and held in the visual-spatial sketchpad component of working memory, as described by Baddeley's Theory of Working Memory. (Kizilcec et al, 2014) Schechinger (2023) elaborates that visualizations can also play a role in reducing cognitive load, which is the amount of mental effort required to process information.

She also says that infographics, which often incorporate visualizations, can decrease cognitive load, allowing viewers to retain the information presented more easily.

It is important to mention that visualizations aren't always beneficial to information retention. While visualizations can aid encoding, they can also lead to cognitive overload if they are too complex or poorly designed. (Kizilcec *et al*, 2014) If a visualization presents too much information or is difficult to understand, it can exceed the capacity of working memory, hindering short-term retention. (Schechinger, 2023)

Notwithstanding, quotes presented here support strong correlations between interactive visualizations and information retention, posing a fertile field to be explored with the addition of juiciness.

## 3. METHODOLOGY

### 3.1. Infographics Development Process

A total of 3 infographics, in 2 versions each, were used for this research. From the 6 total, 1 was available online, considering this the dry version of it, with its juicy version, as well as the other 4 pieces, being developed by the author. The infographics were created with the use of 2 softwares: Adobe Illustrator (source) was used for layout organization and visual assets manipulation, and Construct 3 (source) was the chosen platform to make their interactive versions. The next subtopics provide details about each infographic.

### *3.1.1. Infographic 1: The Daily Routines of Famous Creative People*

This infographic was originally created by Podio.com (2013). Inspired by the book Daily Rituals by Alfred A. Knopf from 2013, this interactive infographic (Figure 1) displays visual information with bar charts that are colour coded, corresponding to different practices (sleep, creative work, day job/admin, food/leisure, exercise and other) from daily routines of famous creative people. Amongst the names presented are Sigmund Freud, Benjamin Franklin and Pablo Picasso, just to name a few. To interact with it, the user can filter the category they want to display, making it visible or invisible by clicking on the labels at the header. Apart from that, beside each name there's an information icon (i) that displays their occupation along with a brief profile description. Despite its interactivity, the only animated feature is the fact that the bar charts fade in or out when a category is turned on or off.

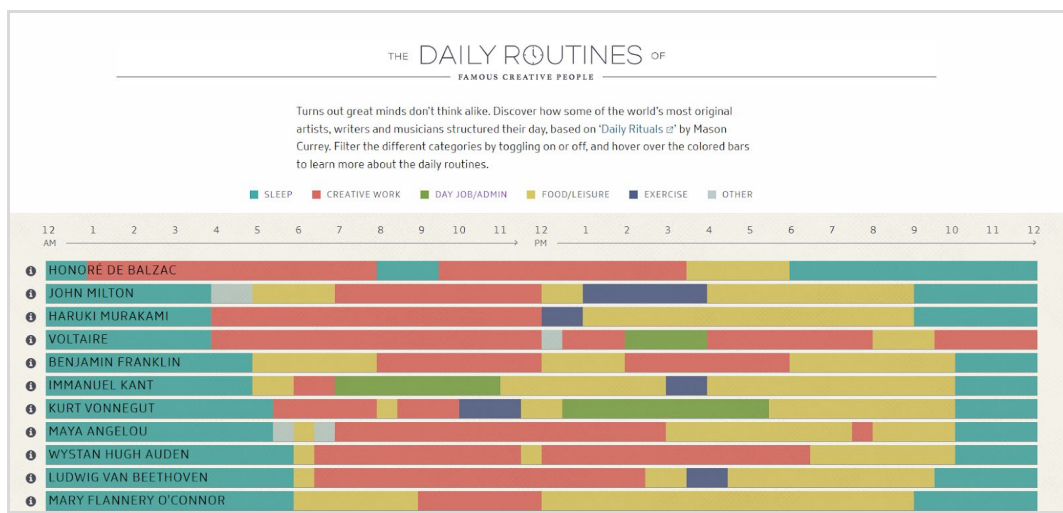

Figure 1. The Daily Routines of Famous Creative People - "Dry" version (Podio.com, 2013)

Some changes were applied to develop a juicier version of the infographic. When loaded, the items subtly fade and move up before landing on their position. The original title had a clock watch merged into the letter "O" from the word Routines, which in the juicy version has spinning pointers. The color squares from the category filters at the top now scale up while hovered with the mouse cursor. If clicked, they spawn some particles. The bar charts now, instead of fading in and out, change their heights with different random timing, the screen shakes, and a sound is played along with it. The information icons (i) also increase while hovered over and the description is displayed with a "scale-up" animation along with a "pop" sound.

### 3.1.2. Infographic 2: Burcalories

Burcalories (Figure 2) is an interactive infographic created by the author, inspired by the SkyBurger game (Nimblebit, 2009). It invites the user to pick items, piling them up to build a burger while checking nutritional values, represented by numbers and bar charts.

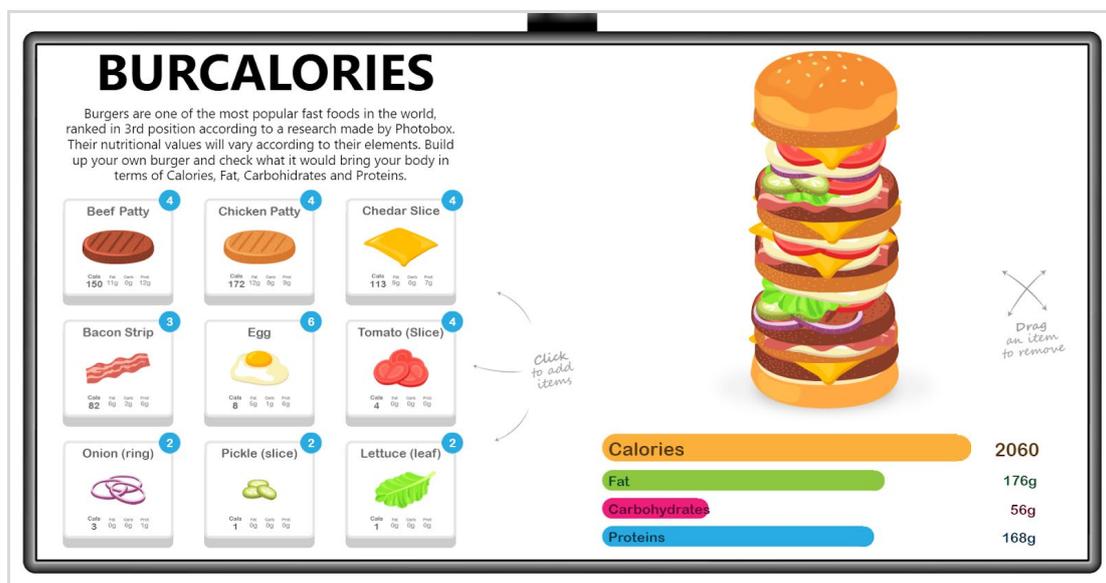

Figure 2. Burcalories - "Juicy" version (Author, 2025)

The dry version instantly loads, revealing all of its components at once. The user can add burger items by clicking on them, making them spawn on the burger. To remove an item, the user needs to click on the desired one inside the burger. No sounds, animations or other visual effects were applied to this version.

The juicy version begins with elements of the infographic fading in. Users start by selecting an item to build the burger. Clicking an item triggers several simultaneous effects: (1) the button simulates a press; (2) the item counter updates and spawns particles; (3) the top bun bounces, adding the item below; (4) bar charts visually adjust based on the item's values; (5) a "pop" sound plays; (6) a "squish" sound plays when items collide. To remove items, users drag and drop them off the burger, reducing counts and updating bar charts. Items behave with physics and collisions, risking multiple items falling and causing the burger to collapse if overloaded.

### 3.1.3. Infographic 3: The Main Chakras

The Main Chakras (Figure 3) is an interactive infographic also created by the author. It provides information about the main 7 chakras which refer to energy points in the human body. According to Kundalini yoga tradition, they are thought to be spinning disks of energy that should stay "open" and aligned, as they correspond to bundles of nerves, major organs, and areas of our energetic body that affect our emotional and physical well-being (Google Arts & Culture, 2022).

Unlike the previous infographics, which feature a single interactive scene, this piece employs two layouts: an introductory overview of the subject and a detailed breakdown of each chakra. The dry version lacks animations, sounds, transitions, or other juicy elements. In the first layout, users interact through "begin" and "next" buttons to progress. The second layout, accessed via an instant frame change, displays a meditating figure with flashing, randomized chakra representations. Clicking a chakra instantly positions it on the figure, changing its color. Once all chakras are set, users can click them again to access additional details displayed on either side of the layout. An information button and a restart button are also available.

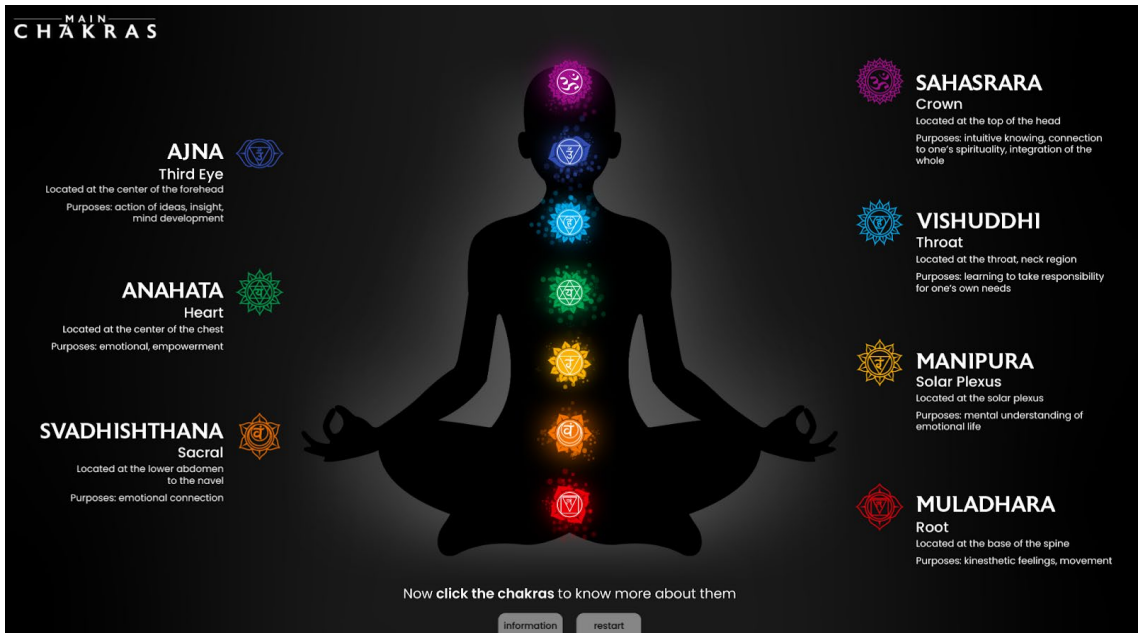

Figure 3. The Main Chakras - "Juicy" version (Author, 2025)

The juicy version emphasizes constant motion and interactivity. Most elements are animated, with some moving and spinning simultaneously. Clickable elements provide multiple forms of feedback. In the initial layout, buttons float up and down, and transitions are smooth with fades. The second layout reveals its components gradually. Unidentified chakras move smoothly across the screen with spinning decorations, leaving particle traces. Users must drag chakras to their body locations, marked by a colored blur. Once placed, clicking a chakra spawns an outlined copy that moves to reveal textual information, which gently oscillates alongside the meditator silhouette, simulating meditative breathing. Most interactions include sound effects.

### 3.2. Participants

Participants for this study come from various backgrounds. They were contacted through multiple channels of contact, mainly via email and WhatsApp, and eventually from word to mouth. Initially, they were informally invited to take part in the research, with a brief explanation. Those who demonstrated interest would share their email address to receive the formal invitation with instructions on how to conduct the test - these will be further explained in the following topic. Data collection took place between March 3rd to 21st, 2025.

### 3.3. Procedures

Once interested individuals provided their email addresses, they received detailed instructions about how to conduct the test, which was divided in 3 steps, happening twice: (1) interact with a dry version of one infographic via the link provided; (2) evaluate the infographic from the User

Engagement Scale (O'Brien *et al*, 2018); and (3) responding to a short quiz about the infographic content. The second part would consist of the same steps, but this time interacting with a different infographic at its juicy version. Figure 4 displays how the distribution was made.

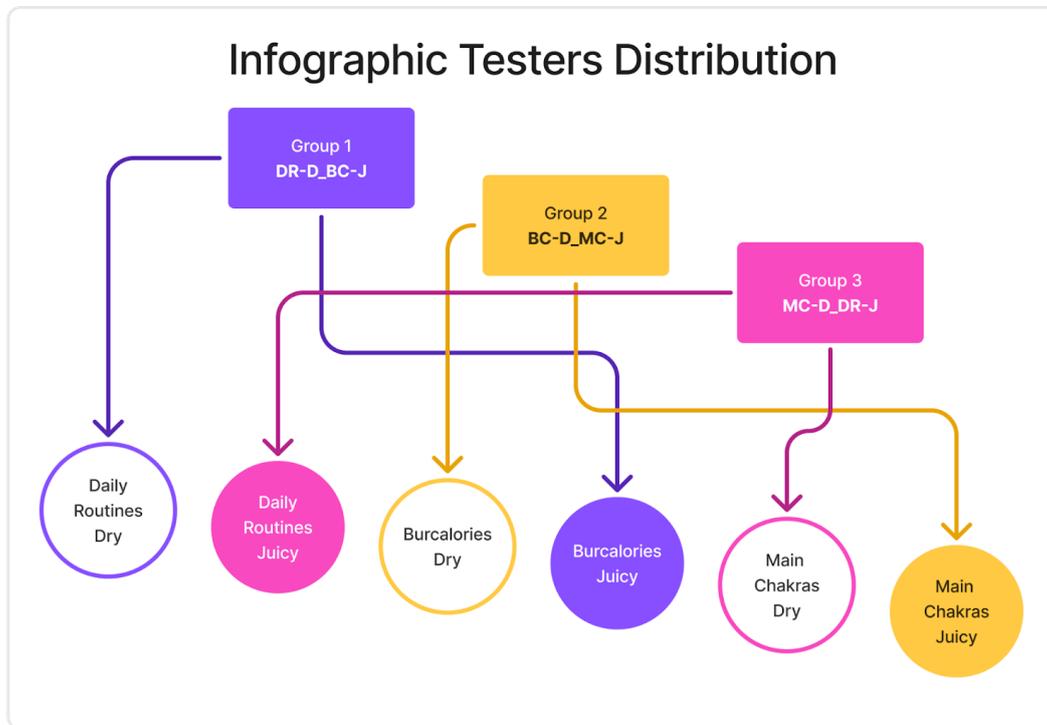

Figure 4. Participants distribution (Author, 2025)

Participants were distributed like that for 2 main reasons: (1) they would have a chance to test one dry and one juicy version of infographics; and (2) they were not testing the same infographic twice so they could be tested for information retention with 2 different pieces, allowing for further comparisons once all the tests were completed. Respondents were assigned with unique identifiers based on the infographics they've interacted with.

The User Engagement Scale (UES) and the information retention for each infographic were organized into Google forms.

### 3.3.1. User Engagement Scale

The original UES, proposed by O'Brien (*et al*, 2018) consisted of 31 items aiming to measure six dimensions of engagement:

- **Focused attention (FA)**, which refers to feeling absorbed in the interaction and losing track of time.
- **Perceived usability (PU)**, related to negative affect experienced as a result of the interaction and the degree of control and effort expended.
- **Aesthetic appeal (AE)**, meaning the attractiveness and visual appeal of the interface.
- **Endurability (EN)** or the overall success of the interaction and users 'willingness to recommend an application to others or engage with it in future.
- **Novelty (NO)**, curiosity and interest in the interactive task.
- **Felt involvement (FI)**, the sense of being "drawn in" and having fun.

A refined version and a shorter form (UES-SF) have also been developed, grouping items from EN, NO and FI into a Reward Factor (RW). The short-form version of the User Engagement Scale was the chosen tool for testing if engagement is affected by Juiciness. Minor wording adjustments were made to align the form with infographics. Table 1 outlines the question organization.

Table 1. Adaptation of the User Engagement Scale Short Form (O'Brien *et al*, 2018)

|  | Strongly disagree | Disagree | Neither agree nor disagree | Agree | Strongly Agree |
|---|---|---|---|---|---|
|  | 1 | 2 | 3 | 4 | 5 |
| FA-S.1 | I lost myself in this experience | | | | |
| FA-S.2 | The time I spent using this infographic just slipped away. | | | | |
| FA-S.3 | I was absorbed in this experience. | | | | |
| PU-S.1 | I felt frustrated while using this infographic. | | | | |
| PU-S.2 | I found this infographic confusing to use | | | | |
| PU-S.3 | Using this infographic was difficult | | | | |
| AE-S.1 | This infographic was attractive | | | | |
| AE-S.2 | This infographic was aesthetically appealing | | | | |
| AE-S.3 | This infographic appealed to my senses | | | | |
| RW-S.1 | Using this infographic was worthwhile | | | | |
| RW-S.2 | My experience was rewarding | | | | |
| RW-S.3 | I felt interested in this experience | | | | |

There was also an optional open ended question asking respondents if they would like to provide any extra feedback regarding interacting with the infographic.

### *3.3.2. Short-Term Information Retention Questions*

To evaluate short-term information retention, participants responded to questions where they were invited to remember their experience with the infographics. According to Kilzicec (*et al*, 2014), a common method to measure short-term retention is to administer recall tests very soon after the presentation of information. For doing so, specific questions were applied for each infographic subject:

- The Daily Routines of Famous Creative People

    *Q: From memory, name as many daily routines categories as you can (open-ended question)*

- Burcalories:

*Q1: Which of the following items has the most Carbohydrates? (multiple-choice question)*

*Q2: Which of the following elements is the most Caloric? (multiple-choice question)*

- The Main Chakras:

   *Q: From memory, name as many chakra body locations as you can. (open-ended question)*

Along with those, for all the infographics users were prompted with the following question: *Q: What is your take away from this infographic? Please elaborate your answer with as much detail as you can.*

The open-ended responses were organized into documents that were later analyzed with Google Notebook LM. The application supported pattern recognition from answers, as well as identifying repeated terms and providing an overview of the responses from each of the 2 variants of the 3 infographics.

The following section discusses the results obtained from the data collected with the forms.

## 4. RESULTS

### 4.1. User Engagement Scale Scores

When comparing the 2 variants of each infographic, juicy versions display a small advantage in terms of overall engagement, as shown in Figure 5.

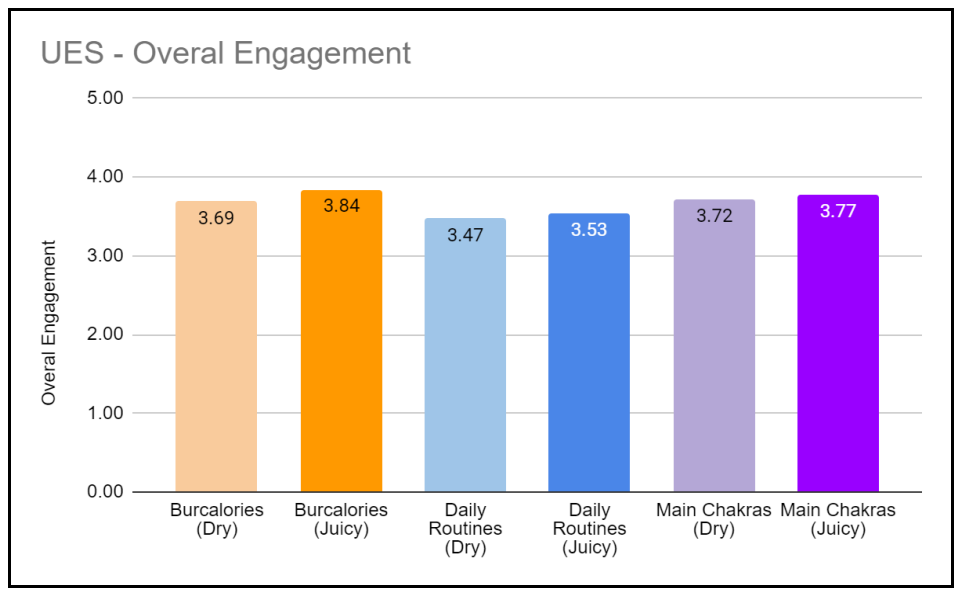

Figure 5. UES Overall Engagement results (Author, 2025)

The juicy version of the Burcalories infographic presented the highest score (3.84 / 5.00), whereas the dry version of The Daily Routines of Famous Creative People has the lowest score (3.47 / 5.00). Still, the score difference between the 2 infographics is only 0.37 points. Burcalories also shows the highest difference comparing its two versions, where the juicy version (3.84) is 0.15 higher than the dry version. The infographic with the smallest difference between the 2 versions was The Main Chakras, with the juicy version being only 0.05 points higher than its dry version.

As previously seen (Figure 4), participants did not access 2 versions of the same infographic. Figure 6 shows the comparison results from the infographics that were accessed within the same test process by participants groups. In this case, one juicy infographic - *The Daily Routines of Famous Creative People* - presented a lower engagement level when compared to the dry infographic - *The Main Chakras* - with a difference of 0.19 points.

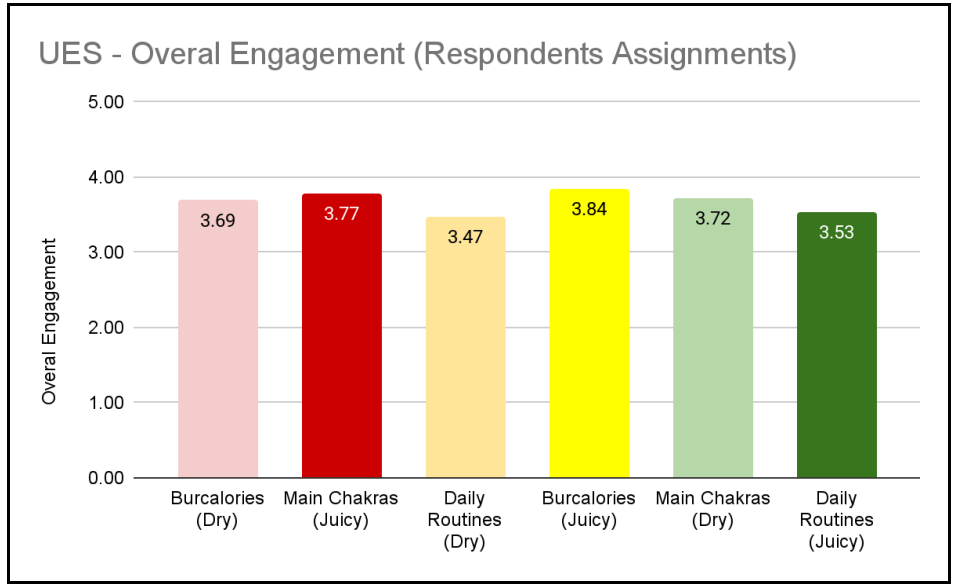

Figure 6. UES Overall Engagement results comparing infographics assigned to the same respondents group. (Author, 2025)

Average scores for each subscale were also calculated. Results (Figure 7) are somewhat similar to Overall Engagement.

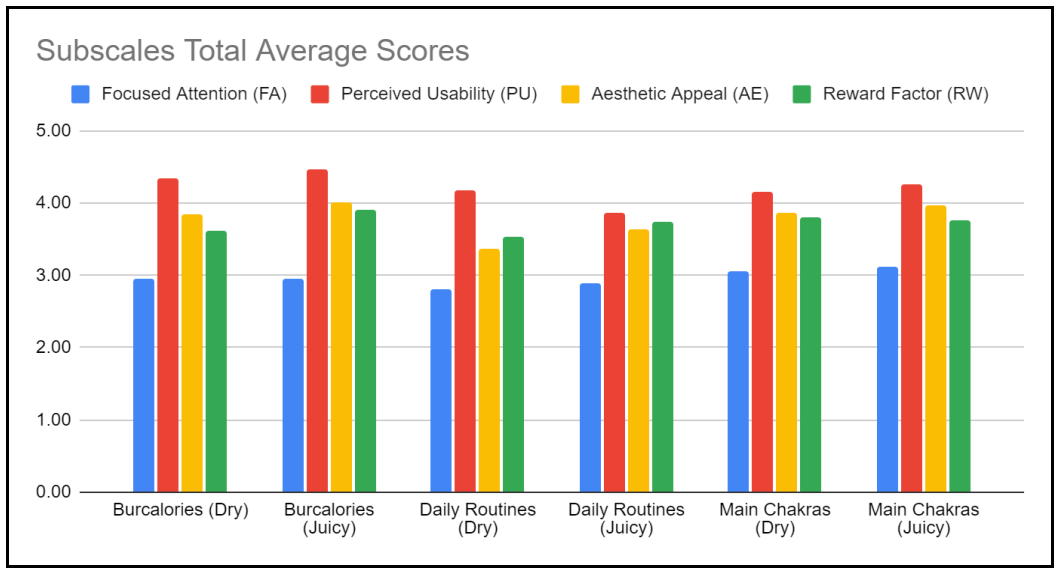

Figure 7. UES Subscales Total Average Scores (Author, 2025)

Subscales results show that Perceived Usability (PU) presents the higher scores amongst the 4 categories, whereas Focused Attention (FA) showed the lowest averages for each infographic.

In two categories, each one from two different infographics, there were examples where dry versions presented higher scores than their juicy counterparts. For Perceived Usability (PU), The Daily Routines of Famous Creative People dry version scored 4.17 versus 3.87 from its juicy version. Reward Factor (RW) was slightly higher for the dry version of The Main Chakras (RW = 3.80) compared to its juicy version (RW = 3.76).

In summary, the total averages do not significantly differ much between dry and juicy versions, nor as well between different infographics.

### 4.2. Feedback and Takeaways

#### *4.2.1. The Daily Routines of Famous Creative People*

Both versions of the infographic were generally found to be interesting and informative, sparking curiosity about the daily habits of famous creative individuals. Respondents appreciated the opportunity to learn about these routines and observe potential patterns or the lack thereof. From the Dry version, one respondent stated, "The theme was unexpected and interesting. I enjoyed it..." Similarly, a respondent from the Juicy version mentioned, "These were really interesting curiosities, I loved learning about them."

The interactive nature of both infographics was also seen as a positive aspect by some users, allowing them to engage with the data in a more meaningful way. The ability to filter categories in the Juicy version was particularly well-received by some. A Dry version user commented, "I really enjoyed the interactivity with the infographic, simple and direct, achieving its goal of making data visualization easier for the user.". Regarding the Juicy version, a user said, "The fact I could filter the categories was really attractive to me."

The infographic served as a reminder that there isn't a single "right" way to be creative or successful, as routines varied significantly among the individuals presented. This was a key insight for many respondents. A Dry version user noted, "That creators have different routines, some having a completely opposite routine from regular day-to-day job routine regular people have." A Juicy version respondent echoed this, stating, "The routine is variable by person, there are no static rules to be creative. It's a matter of finding which routine works better for you."

A significant negative aspect for some users, particularly with the Juicy version, was difficulty in navigating and understanding the infographic. Some found the amount of information overwhelming, or the interaction methods unclear. One Dry version user admitted, "I felt completely lost at first, but it made sense after awhile. However, it has a lot of information, so it was too much for me to absorb it.". A Juicy version respondent stated, "The infographic was difficult to read and to be able to identify each of the data in a simple way, there was a lot of data organized but difficult to differentiate which made it confusing, I could not learn the information, and it was not interactive and that made it a little boring.". Another user also mentioned, "It took me a while to realize that only the Information icon on the left actually contained some interaction and provided extra information. I thought that the bars that defined their daily life were clickable, which made me feel lost on what I was supposed to do.".

Some users in both versions expressed a desire for more context or supporting information, such as the sources of the routine data or potentially pictures of the famous individuals. A Dry version user wondered, "...how do they know these people's routines? Where did this come from? I guess that would have been interesting to add." Another suggested, "It could have a picture of each famous person.".

One specific criticism of the Juicy version was that comparing categories visually was difficult despite the use of different colours. A user commented, "The different colors of the graph was easy to identify the categories but was difficult to compare them. I think with numbers I can visualize more clearly." Furthermore, one user of the Juicy version found the interaction frustrating and difficult to figure out, hindering their ability to learn from the infographic. A respondent said, "It's hard to remember any information on this infographic as I was spending more time trying to figure out how to interact with it. Very frustrating".

*4.2.2. Burcalories*

Both versions of the infographic were generally perceived as interesting and fun ways to learn about the nutritional values of burger ingredients and how they contribute to the overall calorie count and macronutrient breakdown. Respondents from both versions noted that the interactive nature of the tool made it engaging. One comment from a respondent that interacted with the juicy version reads "This infographic was fun! The way the ingredients fall down when you remove them was funny and appealing, like a real-life game or toy." One dry version tester wrote "I found it interesting to visualize the total calories depending on what you included in the burger, and interacting with the ingredients/burger was fun."

Many users found the infographic helpful for understanding the impact of different ingredients on the nutritional profile of a burger. This led to insights about making more informed food choices and being more aware of calorie intake. A Dry version user noted, "This infographic shows how different ingredients in a burger contribute to my nutritional intake in terms of calories, fat, carbohydrates, and proteins. It allows me to customize a burger and see the impact on my daily food intake." A Juicy version user echoed this, stating, "I was able to gather information on how to make healthier food choices."

The visual appeal and ease of use were also highlighted as positive aspects in both versions. For the Dry version, a respondent commented, "The graphic is well-designed, and I appreciate the colors and drawing style. It's very clean and easy to work with...". Regarding the Juicy version, a user said, "It was fun and easy to see how many calories I was adding with each ingredient.".

A significant point of concern across both versions was the lack of nutritional information for the bread buns. Respondents felt this was a crucial omission for accurate calorie and nutrient tracking. A Dry version user stated, "I feel like the bread buns should've also had it's calories/nutrients tracked." A Juicy version respondent also pointed out, "...it felt really weird that the bread was not counting as any nutritional value, that made it feel wrong.".

Another common issue, particularly in the Dry version feedback, was related to the user interface for removing ingredients. Some users found it unintuitive that they had to click directly on the burger to remove items, expecting +/- signs or buttons instead. One Dry version user explained, "I had a bit of difficulty understanding that I should click on the burger directly to remove items - maybe because I am conditioned to the +/- sign near the quantity from other interfaces". Another elaborated, "I also felt these instructions were a little confusing, since I expected that there would be a button to click for removing an item (similar to the button used to add items), instead of directly clicking the item in the burger itself. Clicking the items in the burger to remove them sometimes removed more than one time."

In the Juicy version, some users reported technical issues and interface bugs, such as the carbohydrate count not appearing or issues with items falling and not stopping. One Juicy version user mentioned, "I don't know if it was my mistake, but the carbs never appeared." Another described, "...when I tried to remove the ingredients, the top bread started to fall and come back and it was quite annoying to wait for it to stop falling so I could continue the game (it kept falling maybe 5 times non stop until it stopped)".

In summary, both the Dry and Juicy versions of the Burcalories infographic were valued for their interactive and educational features. The Juicy version offered a more entertaining experience with animations and sounds but encountered technical issues. The Dry version was considered functional yet less engaging, with improvements needed in interface intuitiveness and the inclusion of the bread's nutritional value in both versions.

### *4.2.3. The Main Chakras*

Both versions of the infographic successfully introduced the concept of chakras to many respondents and provided them with some basic knowledge about their existence and potential role in the body. Many users, especially those with little to no prior knowledge, found the experience interesting and informative. From the Dry version, one respondent stated, "I liked to know about the chakras.". Another noted, "I've learned that our body has different chakras and points of energy to be explored.". Similarly, a user of the Juicy version said, "It was interesting and informative. I had no information about chakras and was entrained in playing with the symbols while reading the descriptions about each chakra.". Another echoed this, saying, "Interesting information as I did not have previous knowledge about chakras".

Both infographics also sparked curiosity and motivated some users to want to learn more about the subject. A respondent from the Dry version mentioned, "the interactive experience brought a lot of curiosity to find out more about each element.". Another stated, "this infographic caught my attention. It was interesting to know the Chakras' positions, and know that it is related to different human feelings, actions and health. I will do research on that and bring it to my personal life.". For the Juicy version, a user commented, "It felt very exploratory and built my curiosity - making me wonder what would happen after I drag each symbol to the corresponding body part.", and another said "the visual made me want to know more about the chakras."

Despite the positive aspects, both versions also received criticism. A recurring issue in the Juicy version was the interaction involving moving chakras, which some users found frustrating or distracting. One Juicy version respondent mentioned, "Having to 'chase' the white images to place them was not very fun, in my opinion.", and another stated, "Sometimes I had to chase the little spinning chakras around to catch and move them". Furthermore, the amount of movement in the Juicy version, such as the circular movements while reading, was considered "a bit too much" by some.

The Dry version's initial moving elements also caused some confusion. One user reported, "but at the beginning with all of them moving I did not know what to do.". Additionally, one respondent found the Dry version failed to promote user engagement due to a lack of interest in the subject, a black background, and a lack of visual feedback on clicks, leading to frustration.

Some users in both versions felt that the infographics could have provided more in-depth explanations. A Dry version respondent suggested, "There could have been more explanations about this and more movement, which would have been more illustrative and eye-catching to remain in the memory.". A Juicy version user felt the experience was "really superficial" and another wished to know "more about how to increase the ways I engage with each of these chakras to better find balance."

Finally, some users in the Juicy version found the number of interactions and animations excessive, making the experience feel slow and drawn out. One Juicy version user stated, "I believe that the infographic, while visually appealing, requires too many interactions to access all the information, which seems pointless and makes the experience drag on too long, with considerable time spent just waiting for animations to finish."

## 4.3. Information Retention Results

### *4.3.1. The Daily Routines of Famous Creative People*

For this infographic, participants were asked to recall as many content categories as they could from memory. The 6 correct categories were: Sleep, Creative Work, Day Job/Admin, Food/Leisure, Exercise, and Other. The results (Table 2) provide a comparative overview of how each version may have supported participants' ability to absorb and remember structured information.

Table 2. The Daily Routines of Famous Creative People Infographic: Information Retention Comparison Dry x Juicy

| Metric | Dry Version | Juicy Version |
| --- | --- | --- |
| Average Recall | 4.06 | **4.17** |
| Perfect Recalls (6/6) | 2 | **4** |
| Recalls (5/6) | 3 | **6** |
| Recalls (4/6) | 7 | 3 |
| Recalls (3/6) | 1 | 1 |
| Recalls (2/6) | 1 | 1 |
| Recalls (1/6) | 1 | 2 |
| Most Recalled Category | Sleep (16) | Sleep (17) |

The Juicy version demonstrated a marginally higher average recall (4.17 categories) compared to the Dry version (4.06). Furthermore, the Juicy version yielded more participants achieving full recall of all six categories. Table 2 demonstrates the differences in recall distribution, with the Juicy version showing more respondents recalling five or six categories, suggesting a potential impact of Juicy Design on overall cognitive engagement and memory encoding.

These findings indicate a modest but observable advantage in information retention for the Juicy version of the infographic. While the differences in average recall were not dramatic, the Juicy version supported more complete recall and reduced confusion or misinterpretation.

### *4.3.2. Burcalories*

Answer choices from the 2 multiple choice questions differ between versions. For the first question, which asked users to select the item richer in carbohydrates, the dry version collected 12 correct answers, whereas the juicy version got 7 correct responses (Figure 8).

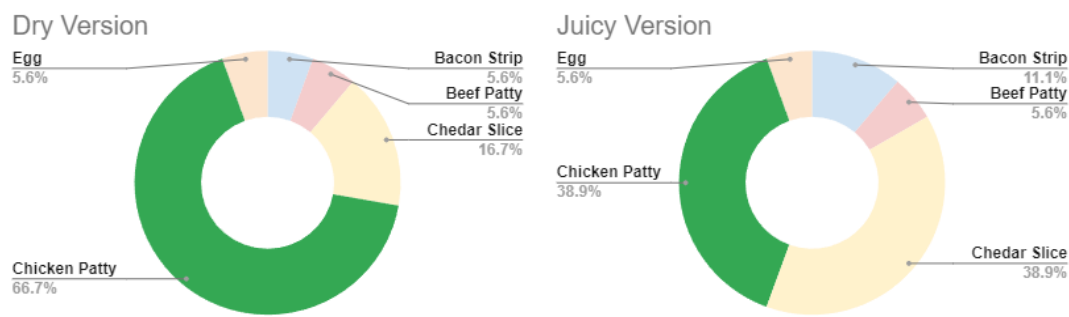

Figure 8. Responses comparison from Burcalories infographics multiple choice question 1 (Author, 2025)

Question 2 inquired about the most caloric item, and the results from each version are even more distinguished: 8 correct answers for the dry version and just 1 correct answer on the juicy version (Figure 9).

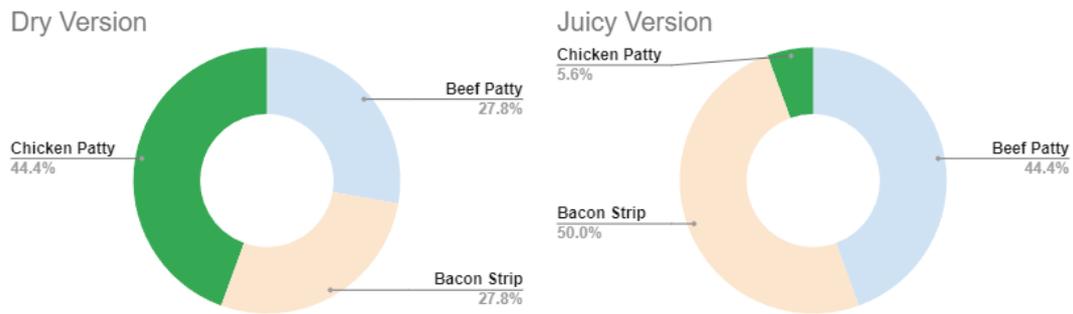

Figure 9. Responses comparison from Burcalories infographics multiple choice question 2 (Author, 2025)

The correct choice for both questions was the "Chicken Patty" item, and its nutritional values were available within the button used to place that item into the burger. At the same time, by adding the Chicken Patty item to the burger, the bar graphs would stretch longer than when other items were placed.

Despite the fact the Dry version collected a higher number of correct answers, in both versions it is unclear if that information was obtained based on the data displayed within the infographic or if the respondent trusted their own senses to select an option.

### 4.3.3. The Main Chakras

Similarly to the analysis conducted for The Daily Routines of Famous Creative People infographics, participants were asked to recall and name as many chakra body locations as they could from memory. The responses were then coded and quantified based on the correct recognition of the seven standard chakra positions. The goal was to evaluate how design differences influence the quantity and distribution of recalled information. Table 3 presents the results.

Table 3. The Main Chakras: Information Retention Comparison Dry x Juicy

| Metric | Dry Version | Juicy Version |
| --- | --- | --- |
| Avg. correct chakra locations recalled | 3.78 | **4.61** |
| Mode (most frequent score) | 6 (4 participants) | **6 (6 participants)** |
| Participants with 0 correct | 3 | **1** |
| Participants with 7 correct | **3** | 2 |
| Participants recalling 5+ chakras | 8 | **10** |

| Most Recalled Chakra | Heart (Anahata) – 14 mentions | Heart (Anahata) – 17 mentions |

The results show a subtle yet meaningful difference in retention outcomes between the two versions of the Chakras infographic. While the Dry version had slightly more top scorers (7 out of 7), the Juicy version produced a broader concentration of high scorers, especially among those recalling 5 or 6 chakras. These findings suggest that a Juicy design may facilitate better average recall and reduce the number of participants who forget most of the content. The richer visual cues and interactions of the Juicy version appear to offer more anchors for memory, reinforcing the hypothesis that juiciness positively influences information retention in interactive infographics.

## 5. DISCUSSION

Feedback from the two interactive infographic versions highlighted a tension between engagement and cognitive load. The Juicy versions were praised for their entertaining and emotionally engaging features, such as animations and sound effects, but some users reported frustration with excessive or unclear interactions. For example, in the Burcalories infographic, falling burger ingredients were fun for some but caused delays in processing information. In the Chakras infographic, spinning chakras distracted users, making them harder to move and follow, which hindered educational content absorption. Conversely, the Dry versions offered simpler navigation and clearer data visualization but were perceived as less engaging and occasionally superficial.

Results from the User Engagement Scale (UES) indicate minimal differences in Overall Engagement and subscale scores between the Dry and Juicy versions across the infographics. While Juicy versions showed a slight advantage in overall engagement, Dry versions often outperformed in categories like Perceived Usability and Reward Factor. These findings suggest that while juicy elements, such as interactivity and animation, can enhance emotional appeal, they are not always essential for user satisfaction or information retention. This highlights the importance of prioritizing clarity and usability over novelty or visual complexity, especially for unfamiliar subject matter.

Interactivity, such as filtering, moving elements, and visualizing data dynamically, was positively received in both Juicy and Dry versions. In the Daily Routines and Burcalories infographics, users valued the ability to customize interactions, like filtering categories or observing how ingredients affected nutritional values, fostering a sense of agency and personalization. However, in the Juicy versions, some interactions were less intuitive, causing confusion and frustration. For example, in the Chakras infographic, users often struggled to move chakras to their corresponding body parts, detracting from the experience. These findings highlight the need for interactive features to be both engaging and intuitive, ensuring alignment with user expectations.

The Juicy elements, such as animations and sound effects, significantly enhanced the emotional appeal of interactive infographics. For instance, users appreciated the "pop" sounds in the Burcalories infographic, noting they added fun and engagement. These features fostered positive emotional responses, making the learning process more enjoyable. However, while they increased engagement, they did not always improve information retention. In the Juicy version of the Daily Routines infographic, some users felt overwhelmed by the data and struggled to navigate the interactive layers. This underscores the need for emotional design elements to align with and support learning objectives rather than function as purely decorative features.

Regarding information retention, the Juicy versions of the infographics generally showed a modest improvement, particularly in average recall and the number of participants achieving complete recall. The Juicy design's interactive elements and richer visual cues appear to support better memory encoding and retrieval, leading to slightly improved recall rates across different types of content. However, the differences between Dry and Juicy versions were not always dramatic, suggesting that while interactivity and visual richness enhance retention, the content type and cognitive engagement also play significant roles in determining the success of information retention in interactive infographics.

## 6. CONCLUSION

The present study proposed a comparison between juicy and dry versions of 3 infographics to verify how they would perform in terms of user engagement and information retention levels amongst users. Even though with minor differences, results demonstrate that both proposed hypotheses are partially achieved. *H1: Interactive infographics with juicy design elements will result in higher user engagement scores than dry/regular infographics* was identified for all the infographics, while *H2: Users interacting with juicy infographics will exhibit better short-term information rates retention compared to those interacting with dry/regular infographics* applies to The Daily Routines of Famous Creative People and The Main Chakras.

As previously demonstrated by other research (Durmanova, 2022; Singhal and Schneider, 2021; Kao, 2020; Hicks, 2020), this study findings suggest that juiciness can significantly enhance user experience by making interactions more enjoyable, immersive, and aesthetically pleasing. However, the key lies in thoughtful design and moderation. Designers need to carefully consider the context and the amount of "juiciness" to apply, as both insufficient and excessive feedback can negatively impact the user's overall experience. When implemented well, "juiciness" can be a powerful tool for creating positive and engaging user experiences across a wide range of applications.

Future studies could incorporate more robust software for visual content creation, such as Unity and Unreal Engine, offering the possibility to create 3D assets and more complex coding. It is also suggested that future studies compare different versions of the same infographic, with various levels of juiciness, with the same testers trying all of those versions, in order to gather impressions from different versions with the same subject and content. That could lead to opportunities to identify optimal levels of juicy design for interactive infographics.

## REFERENCES

Afify, M. K. (2018). The Effect of the Difference Between Infographic Designing Types (Static vs Animated) on Developing Visual Learning Designing Skills and Recognition of its Elements and Principles. International Journal of Emerging Technologies in Learning (iJET), 13(09), 204. https://doi.org/10.3991/ijet.v13i09.8541

Atanasov, S. (2013). Juiciness: Exploring and designing around experience of feedback in video games. In K3, MAH, Malmö, Sweden [Thesis-project].

Burnett, E., Holt, J., Borron, A., & Wojdynski, B. (2019). Interactive infographics' effect on elaboration in agricultural communication. Journal of Applied Communications, 103(3). https://doi.org/10.4148/1051-0834.2272

Campos, B. (2025). Assessing juicy elements in interactive infographics. Revista De Comunicación De La SEECI. https://doi.org/10.15198/seeci.2025.58.e909

Coelho, D., & Mueller, K. (2020). Infomages: Embedding Data into Thematic Images. Computer Graphics Forum, 39(3), 593–606. https://doi.org/10.1111/cgf.14004

Costa Pinto, J. & University of Santiago de Compostela. (2017). The relevance of digital infographics in online newspapers. European Scientific Journal, 428–430.

Dehghani, M., Mohammadhasani, N., Ghalevandi, M. H., & Azimi, E. (2020). Applying AR-based infographics to enhance learning of the heart and cardiac cycle in biology class. Interactive Learning Environments, 31(1), 185–200. https://doi.org/10.1080/10494820.2020.1765394

Diakopoulos, N., Kivran-Swaine, F., Naaman, M., & School of Communication and Information, Rutgers University. (2011). Playable Data: Characterizing the design space of game-y infographics. In CHI 2011 [Conference-proceeding].

Durmanova, K. (2022, December 21). The Effects of Juicy Game Design on Exergames. http://hdl.handle.net/10012/18980

Fabre, É., Seaborn, K., Verhulst, A. A., Itoh, Y., & Rekimoto, J. (2024). Juicy Text: Onomatopoeia and Semantic Text Effects for Juicy Player Experiences. ICMI'24 - 26th ACM International Conference on Multimodal Interaction, 144–153. https://doi.org/10.1145/3678957.3685755

Fischer, Laura Morgan; Elizabeth Schroeder; Gibson, Courtney; McCord, Amber; and Orton, Ginger (2023) "An Experimental Study Investigating the Type of Data Visualizations Used in Infographics on Participant Recall and Information Recognition," Journal of Applied Communications: Vol. 107: Iss. 3. https://doi.org/10.4148/1051-0834.2489

Ge, T., Zhao, Y., Lee, B., Ren, D., Chen, B., & Wang, Y. (2020). Canis: a High-Level language for Data-Driven Chart Animations. Computer Graphics Forum, 39(3), 607–617. https://doi.org/10.1111/cgf.14005

Gonzalez, L. S. (2018). Aspects of a Literacy of Infographics: Results from an Empirical-Qualitative Study. https://escholarship.org/uc/item/8kp4n7f4

Google NotebookLM | Note Taking & Research Assistant powered by AI. (access in March 2025). Google NotebookLM. https://notebooklm.google/

Greussing, E., & Boomgaarden, H. G. (2021). Promises and pitfalls: Taking a closer look at how interactive infographics affect learning from news. International Journal of Communication, 15, 22. https://ijoc.org/index.php/ijoc/article/view/15419

Hicks, K., The University of Lincoln Corporate Guidelines, & School of Computer Science, College of Science, University of Lincoln. (2020). Juicy Game Design: Exploring the Impact of "juiciness" on the Player Experience (By K. Gerling, S. Björk, M. Flintham, & K. Bachour) [Thesis].

Johansen, M., Cook, M., IT University of Copenhagen, & Queen Mary University of London. (2021). Challenges in Generating Juice Effects for Automatically Designed Games. In Proceedings of the Seventeenth AAAI Conference on Artificial Intelligence and Interactive Digital Entertainment.

Kao, D. (2020). The effects of "juiciness" in an action RPG. Entertainment Computing, 34, 100359. https://doi.org/10.1016/j.entcom.2020.100359

Kao, D., Ballou, N., Gerling, K., Breitsohl, H., & Deterding, S. (2024). How does Juicy Game Feedback Motivate? Testing Curiosity, Competence, and Effectance. CHI '24: Proceedings of the 2024 CHI Conference on Human Factors in Computing Systems, 1–16. https://doi.org/10.1145/3613904.3642656

Kizilcec, R. F., Papadopoulos, K., & Sritanyaratana, L. (2014). Showing face in video instruction. CHI 2014, 2095–2102. https://doi.org/10.1145/2556288.2557207

Kucic, M. (2005). How to prototype a game in under 7 days. https://www.gamedeveloper.com/game-platforms/how-to-prototype-a-game-in-under-7-days

Lalmas, M., O'Brien, H., & Yom-Tov, E. (2014). Measuring user engagement. Synthesis Lectures on Information Concepts, Retrieval, and Services, 6(4), 1–132. https://doi.org/10.2200/s00605ed1v01y201410icr038

Luz, A., Marcher, F., Nacke, L. E., & Vogel, D. (2024). Encouraging Disengagement: Using Eye Tracking to Examine Attention with Different Levels of Juicy Design. 17th International Conference on Advanced Visual Interfaces, 1–5. https://doi.org/10.1145/3656650.3656694

O'Brien, H. L., Cairns, P., & Hall, M. (2018). A practical approach to measuring user engagement with the refined user engagement scale (UES) and new UES short form. International Journal of Human-Computer Studies, 112, 28–39. https://doi.org/10.1016/j.ijhcs.2018.01.004

O'Day, D. H. (2007). The Value of Animations in Biology Teaching: A Study of Long-Term Memory Retention. CBE—Life Sciences Education, 6(3), 217–223. https://doi.org/10.1187/cbe.07-01-0002

Sl, U. T. (n.d.). Sky Burger for Android - Download the APK from Uptodown. Uptodown. https://sky-burger.en.uptodown.com/android

Schechinger, K. (2023). What do you remember? An analysis of information retention and recall through data visualization use in infographics. In C. Gibson, L. Fischer, A. McCord, & M. Sheridan (Eds.), Agricultural Communications.

Singhal, T., & Schneider, O. (2021). Juicy Haptic Design: Vibrotactile embellishments can improve player experience in games. In CHI - Computer Human Interaction (pp. 1–11). https://doi.org/10.1145/3411764.3445463

The 7 Chakras and Kundalini Energy - Google Arts & Culture. (n.d.). Google Arts & Culture. https://artsandculture.google.com/story/the-7-chakras-and-kundalini-energy-the-yoga-institute/OwUheMZOAtNOLQ?hl=en

The daily routines of famous creative people. (n.d.). Podio.com. https://web.archive.org/web/20231021065401/https://podio.com/site/creative-routines

Won, J. (2018). Interactive Infographics and Delivery of information : The value assessment of infographics and their relation to user response. Archives of Design Research, 31(1), 57–69. https://doi.org/10.15187/adr.2018.02.31.1.57

Yildirim, S. (2024). APPROACHES OF DESIGNERS IN THE DEVELOPED EDUCATIONAL PURPOSES OF INFOGRAPHICS' DESIGN PROCESSES. Zenodo. https://doi.org/10.5281/zenodo.231283

Zeglen, E., Rosendale, J., National Center for Educational Statistics, Babson Survey Research group, Cavanagh, Thompson, Adams, DeFleur, Rosendale, Lindsey, Shroyer, Pashler, Mozer, Betts, Bal, Coll, Rochera, Gispert, Hattie, . . . Meneses. (2018). Increasing online information retention: Analysing the effects of visual hints and feedback in educational games [Journal-article].

Zwinger, S., & Zeiller, M. (2016). Interactive Infographics in German Online Newspapers. In University of Applied Sciences Burgenland & University of Applied Sciences Burgenland, University of Applied Sciences Burgenland.

Zwinger, S., Langer, J., & Zeiller, M. (2017). Acceptance and Usability of Interactive Infographics in Online Newspapers. 2017 21st International Conference Information Visualisation (IV). https://doi.org/10.1109/iv.2017.65